\documentclass[12pt]{article}

\usepackage{amsthm,amsmath,natbib}
\RequirePackage[colorlinks,citecolor=blue,urlcolor=blue]{hyperref}
\usepackage{epsfig,multicol,multirow}
\usepackage{geometry}
\renewcommand{\baselinestretch}{1}

\def\se{\hbox{SE}}

\def\var{\hbox{Var}}
\def\cov{\hbox{Cov}}

\def\max{\hbox{max}}

\newcommand{\U}{\mathcal{U}}
\newcommand{\bI}{\mbox{$\boldsymbol{I}$}}

\begin{document}
\title{Decomposing the Quantile Ratio Index\\ with applications to
Australian income and wealth data}
\author{Luke A. Prendergast\thanks{luke.prendergast@latrobe.edu.au} and Robert G. Staudte\thanks{Corresponding author: r.staudte@latrobe.edu.au}\\
Department of Mathematics and Statistics, La Trobe University\\
Melbourne, Victoria, Australia, 3086}
\date{29 November, 2017}
\maketitle

\begin{abstract} The quantile ratio index introduced by Prendergast and Staudte 2017 is a simple and effective measure of relative inequality for income data that is resistant to outliers. It measures the average relative distance of a randomly chosen income from its symmetric quantile. Another useful property of this index is investigated here:  given a partition of the income distribution into a union of sets of symmetric quantiles, one can find the conditional inequality for each set as
measured by the quantile ratio index and readily
combine them in a weighted average to obtain the index for the entire population. When applied to data for various
years, one can track how these contributions to inequality vary over time, as illustrated here for Australian  income and wealth data.
\end{abstract}


\clearpage
\newpage

\section{Introduction}\label{sec:intro} It is desirable to break down a measure of inequality for a population of incomes into contributions to inequality from sub-populations.  A natural partition of a population for the quantile ratio index (QRI) of \cite{prst-2017b} is provided  by unions of symmetric quantiles.  It allows one to determine how inequality in the middle half, for example, affects the QRI compared to how inequality between the smallest and largest quartiles does.

\subsection{Background}\label{sec:background}

When economists write about \lq decomposability\rq\  of inequality indices they usually mean an ANOVA type breakdown of the inequality index between and within sub-populations,  \citep[and references therein]{bourg-1979,shor-1980,zheng-2007,cowell-2011,subr-2011}. The QRI does not appear to satisfy such a decomposition, but rather lends itself to partitions of unions of symmetric quantiles which are also of interest.

Such partitions arise quite naturally in discussions of inequality. For example, if one defines the \lq middle income class\rq\ as those having incomes lying between the first and third quartile, then given observational data, one can examine how inequality in this middle class varies over time. Further, how does this inequality contribute to the overall population inequality, which may be changing differently over time? Such questions can be answered with the QRI.

As another example, the popular P90/P10 ratio of percentiles is commonly used to compare large with small incomes,
see \cite{burk-2009}, for example. However, such ratios are known to have large standard errors \cite{prst-2017a}
even for large sample sizes.  Moreover a single ratio of quantiles has less information about the rich/poor
ratio than can be obtained using the QRI. Noting that P90 is the median of the upper quintile and P10 is the median
of the lower quintile, here we apply the QRI to the union of the largest 20\% of incomes with the smallest 20\%, and also see how this contributes to the overall QRI for the entire population, as summarized in the weighted average
(\ref{eqn:wtedmean}). By estimating these components of the QRI for data in different years, we can see which
portions of the population are changing over time, as we illustrate with Australian Bureau of Statistics income and wealth data in
Section~\ref{sec:applics}.

To make these statements precise, we next formally introduce
the QRI.
Let $F$ be the cumulative distribution function (cdf) describing a population of non-negative incomes with possible positive mass on zero $F(0)<1/2$, and define
the quantile function of $F$ by  $Q(p)= \inf \{x: F(x)\geq p\}$, $0\leq p<1$.  Further define $Q(1)= \lim _{p\to 1}Q(p)$,
which equals $+\infty $ if $F$ has infinite support. Often we write $x_p$ for $Q(p)$.
Following \cite{prst-2017b}, define the {\em ratio of symmetric quantiles} by $R(p)= Q(p/2)/Q(1-p/2)=x_{p/2}/x_{1-p/2}$ for $0\leq p \leq 1.$  A plot of $R(p)$ against $p$ shows how the typical
(median) income of those with the lowest 100$p$\% incomes, divided by typical (median) income of those with the highest 100$p$\% incomes varies with $p$.

Relative inequality in the population of incomes is measured by the  {\em quantile ratio index}, the area above  $R(p)$ and less than the horizontal line at one:\quad  $I= \int _0^1\{1-R(p)\}\,dp$. Each of the $(1-R(p))$s is itself a measure of relative inequality; for example, $1-R(0.2)=1-x_{0.1}/x_{0.9}$ is the ratio of percentiles P90/P10, after transformation to the unit interval so that larger values indicate more relative
inequality. Thus $I$ is a simple average of these relative inequality measures, one for each $p$, while most such measures, including the Gini index, are a ratio of two measures, concentration and scale.  It is shown in
 \cite{prst-2017b} that $I=I(F)$ has a bounded influence function which explains the good robustness properties
of its estimator $\widehat I,$ defined below in Section~\ref{sec:inference}.
For extensive comparison of the Gini index with $\hat I$ and other outlier resistant measures of relative inequalty, see \cite{prst-2016b, prst-2017b}.

\subsection{Derivation of the decomposition and examples}\label{sec:defns}

Partition the unit interval into symmetric unions of intervals as follows: \ given  $K\geq 1$ and
 $0=p_0<p_1< \dots <p_{K-1}< p_K=1/2$ define $A_k=[p_{k-1},p_k)\cup (1-p_k,1-p_{k-1}]$ for $k=1,\dots, K-1$ and let the last $A_K=[p_{K-1},1-p_{K-1}]$. For our purposes it is useful to think of $A_K$ as essentially
the union of two intervals $[p_{K-1},1/2)$ and $(1/2,1-p_{K-1}],$ the central point 1/2 playing a trivial role in what follows.
 We call $\{A_1,\dots , A_K\}$ a {\em symmetric $K$-partition} of [0,1].
When $K=2$ for any $0<p_1<1/2$  the symmetric 2-partition consists of  $A_1=[0,p_1)\cup (1-p_1,1]$ and $A_2=[p_1,1-p_1]$; and in particular for $p_1=1/4$  we obtain the {\em quartile partition}: the set $A_1$ describes the outer quartiles while $A_2$ the inner quartiles or central half. Another example is the  {\em quintile partition} obtained by taking $K=3$ and $p_1=0.2$, $p_2=0.4$. This symmetric 3-partition consists of $A_1=[0,0.2)\cup (0.8,1.0]$, $A_2=[0.2,0.4)\cup (0.6,0.8]$ and $A_3=[0.4,0.6].$

Next we derive the decomposition of $I$ into a weighted sum of inequality contributions in the partitioning
of $F$ inherited from a symmetric $K$-partition. To this end, when $X$ has cdf $F$, write $X\sim F$; and denote $U\sim \U [0,1]$ for $U$ uniformly distributed on [0,1].
Fix $k$ and let $Q\{A_k\}$ denote the image of $A_k$ under $Q=F^{-1}$.
Also introduce $w_k= \Pr (X\in Q\{A_k\})=\Pr (U\in A_k)=2(p_k-p_{k-1}).$
For simplicity of notation in the coming paragraphs,
temporarily let  $a=p_{k-1}$ and $b=p_{k}$, so $A_k=[a,b)\cup (1-b,1-a]$ and $c=2(b-a)=w_k$. The  conditional distribution of $F$ given $X \in Q\{A_k\}$ has cdf $F_k$ given by:
\[F_k(x) = \left\{
\begin{array}{lllll}
\frac {F(x)-a}{c},                   & \hbox{ for }\qquad\qquad    Q(a) \leq  x < Q(b)       \ ; \\
\frac {1}{2},                        & \hbox{ for }\qquad\qquad    Q(b) \leq  x < Q(1-b)     \ ; \\
\frac {1}{2}+\frac {F(x)-(1-b)}{c},  & \hbox{ for }\qquad  Q(1-b) \leq  x < Q(1-a) \ .
 \end{array} \right.
\]
The quantile function $Q_k$ of $F_k$ can be obtained by solving for $x=Q_k(u)$ in the above expression to obtain:
\[Q_k(u) = \left\{
\begin{array}{ll}
Q(a+cu),          & \hbox{ for }\qquad\quad   0  <    u    \leq  1/2       \ ; \\
Q(cu+1+a-2b),& \hbox{ for }\qquad       1/2  <  u < 1 \ .
 \end{array} \right.
\]
Given $X \in Q\{A_k\}$, the {\em conditional quantile inequality curve} is
defined for $0<p<1 $ by
\begin{equation}\label{eqn:Rk}
R_k(p) = \frac{Q_k(p/2)}{Q_k(1-p/2)}=\frac{Q(a+c\,p/2)}{Q(1-a-c\,p/2)}~.
\end{equation}
Given $X \in Q\{A_k\}$, the {\em conditional QRI} is denoted $I_k$ and determined by
\begin{equation}\label{eqn:Ik}
1-I_k= \int_0^1R_k(p)\;dp  =\int _0 ^1 \frac{Q(a+c\,p/2)}{Q(1-a-c\,p/2)}\,dp=\frac{1}{c} \int_{2a}^{2b}R(u)\,du ~,
\end{equation}
where we have made the change of variable $u=2a+cp.$
Rewriting $1-I_k$ in terms of our earlier notation $a=p_{k-1}$, $b=p_{k}$ and $c=2(b-a)=w_k,$
multiplying both sides of equation (\ref{eqn:Ik}) by $w_k$ and summing over $k$ leads to the weighted average:
\begin{equation}\label{eqn:wtedmean}
I= \sum_kw_kI_k  ~.
\end{equation}
It is evident that $I_1\geq \dots \geq I_K$ because $Q$ is non-decreasing.
Moreover, we can interpret each $I_k$ as the conditional inequality, given $X\in Q\{A_k\}$.
The product $w_kI_k$ gives the amount that the $k$th partition member contributes to $I$.

\marginpar{Figure 1 here.}

\subsection{Symmetric $K$-partitions and inequality decompositions for the standard lognormal
distribution}

 For $F$ the standard lognormal distribution and the {\em quintile partition} defined by $K=3$ with $p_1=0.2$, $p_2=0.4$, the graphs of $R$ and $R_1,R_2,R_3$ are shown in Figure~\ref{fig1}. The shaded areas above the inequality curves and below the horizontal lines at one are respectively $I=0.6638$, $I_1= 0.9171$, $I_2= 0.6352$ and $I_3= 0.2144.$
 By (\ref{eqn:wtedmean}) $I= 0.6638=0.4\times 0.9171+0.4\times 0.6352+
0.2\times 0.2144.$  For this last partition $I_1$ and $I_2$ each contribute 40\% to the overall index $I$
while the middle group contributes only 20\%.   (If one desires all three partition members to contribute
equally to $I$, one needs $p_1=1/6$ and $p_2=1/3.$ Then $I$ is the simple average of  $I_1=0.9334$, $I_2=0.7325$ and $I_3= 0.3254.$)

Next consider the {\em decile partition} determined by $p_k=k/10$, $k=1,2,3,4.$ It yields the decomposition
$I=(I_1+I_2+\dots +I_5)/5= (0.9619+0.8723+0.7376+0.5327+0.2144)/5.$ From this decomposition, one can recover
results for the coarser quintile partition  obtained at the start of this example:
\[I= 0.4\,\left\{\frac{0.9619+0.8723}{2}\right\}+0.4\,\left\{\frac{0.7376+0.5327}{2}\right\}+0.2\,\{0.2144\}~.\]

\subsection{More examples of decompositions of $I$}\label{sec:exsI}
In general the calculation of $\int _0^rR(p)\,dp$ requires numerical integration, but for the lognormal
distribution  closed form expressions for the components $I_k$ of $I$ are obtainable.
The quantile function of the lognormal distribution with parameters $\mu $, $\sigma $ on the log-scale is
$Q_{\mu,\sigma}(p)=\exp \{\mu +z_p\sigma \}$, where $z_p =\Phi ^{-1}(p)$ is the $p$-quantile of the standard normal distribution having cdf $\Phi $.
Therefore $R_\sigma (p)=\exp \{2\sigma z_{p/2}\}$. Using elementary calculations relegated to Appendix~\ref{app1}, one can
show \begin{equation}\label{eqn:Rkindef}
\int _0^rR_\sigma (p)\,dp=2\exp \{2\sigma ^2\}\,\Phi \left \{\Phi ^{-1}\left (\frac{r}{2}\right )-2\sigma \right \}~.
\end{equation} It follows immediately that $I=I(\sigma ) =1-2\exp \{2\sigma ^2\}\,\Phi (-2\sigma ).$
This QRI is monotone increasing from 0 to 1 as the shape parameter $\sigma $ increases from 0 to $\infty.$

We can also find closed form expressions for the ingredients in the decomposition (\ref{eqn:wtedmean}) using
(\ref{eqn:Rkindef}). For a $K$-partition $A_k$ defined by $0=p_0<p_1< \dots <p_{K-1}< p_K=1/2$
we have weights $w_k=2(p_k-p_{k-1})$ and $I_k$ determined by
 \begin{equation}\label{eqn:Iklnorm}
  1- I_k(\sigma )= \frac{1}{w_k}\int _{2p_{k-1}}^{2p_k}R_\sigma (p)\,dp =
   \frac{2\exp \{2\sigma ^2\}}{w_k} \left \{\Phi (z_{p_{k}}-2\sigma )-
       \Phi (z_{p_{k-1}}-2\sigma )\right \}    ~.
\end{equation}

In the top left of Figure~\ref{fig2} are shown the graphs of $I(\sigma )$, $I_1(\sigma )/2$ and $I_2(\sigma )/2$ for the quartile partition.  The inequality $I_1(\sigma )$ rises rapidly to 1 for $\sigma <2$. Therefore for
$\sigma >2$ nearly all the change in the index $I(\sigma)$ for the lognormal is due to that in the central half.

 Quartile partition inequality graphs for the Type II Pareto distribution with shape
parameter $a$ are quite different, see the top right of Figure~\ref{fig2}. In this case $I=I(a)$ is monotone decreasing in $a$ from a high of 1 to its asymptotote value of 0.7016.
The contribution $I_1(a)/2$, inequality due to the outer quartiles
is almost constant, decreasing from a high of 0.5 to a low of $0.4615$ as $a$ increases without  bound. The contribution of inequality in the middle half of the population $I_2(a)/2$ descnds from 0.5 to a low of 0.2401 with
increasing $a$. Thus nearly all the change in inequality comes from the central half of the population.  The asymptotic
values are those  belonging to the exponential distribution, which is the limit of the Pareto$(a)$ distributions as
$a\to \infty .$

\marginpar{Figure 2 here.}

The lower plots in Figures~\ref{fig2} exhibit different behavior from the previous plots, with $I$ descending from 1 to 0 as the shape parameter increases. The contributions to inequality of the inner and outer quartiles in each case start at 0.5 and descend slowly
to 0.  For the symmetric Beta$(\alpha ,\alpha )$ family, the graphs (not shown) are similar in shape to those for
the Gamma$(\alpha )$ family, but they descend to 0 faster as $\alpha $ increases.

\subsubsection*{Equi-$K$-partitions for large $K$.}
In practice we are usually concerned only with small $K$-partitions. But what happens to the decomposition (\ref{eqn:wtedmean}) if we fix $F$ and take an {\em equi-$K$-partition} defined
by $p_k=k/(2K),$ for $k=0,1,2,\dots ,K$, and let $K\to \infty $?
There are then $K$ partition members with equal weights $w_k=2(p_k-p_{k-1})=1/K.$
The inequality in the $k$th subpopulation $F_k$ contributes $w_kI_k=I_k/K$ to the sum in (\ref{eqn:wtedmean}); and, for $k/K\to p$ \[1-I_k= \frac{1}{w_k} \int_{2p_{k-1}}^{2p_{k}}R(u)\,du=K\int_{(k-1)/K}^{k/K}R(u)\,du \to R(p)~.\]
Thus for a large equi-$K$-partition with $k/K \to p$, the conditional inequality $I_k$ is approximately $1-R(p).$

\subsection{Summary of results to follow}\label{sec:summary}

In  the next Section~\ref{sec:inference} we explain how to find asymptotic standard errors and confidence intervals for the  $\widehat I_k$s, and confirm by simulation studies the reliability of coverage probabilities for them for a wide range of possible income distributions.
Applications of the symmetric decomposition theory and methodology to Australian income and wealth data are in Section~\ref{sec:applics}. Details for two web-based applications that we have developed for readers to further study the IQR and to estimate the IQR for their own data are found in Section~\ref{sec:shiny}.  Further applications and extensions are suggested in Section~\ref{sec:conclusion}.

\section{Inference for QRI component estimates}\label{sec:inference}

In this section we obtain large sample confidence intervals for the $I_k$s using methodology
of \citet[Section 3.2]{prst-2017b} to find such intervals for $I$.  The basic idea there was to
choose a positive integer $J$, define a grid $p_j=(j - 1/2)/J$, $j=1,\dots, J$ on the unit interval, and then
estimate $I=\int _0^1(1-R(p))\,dp$ by the average $\widehat I^{(J)}=\{\sum _{j=1}^J[1 - \widehat{R}(p_{j})\}/J$ where
$\widehat{R}(p_{j})$ is an estimate of the quantile ratio $R(p_j) = Q(p_j/2)/Q(1-p_j/2).$ Using standard
results for the asymptotic normality and covariance structure of sample quantiles, nominal 100$(1-\alpha)$\% confidence intervals
for $I$ of the form $\widehat{I}^{(J)} \pm z_{1-\alpha/2}\{\widehat{\var}(\widehat{I}^{(J)})\}^{1/2}$ are obtained.
The required quantile function estimates  are the continuous Type 8 estimates recommended by \cite{hynd-1996} and
the quantile density estimates are those developed by \cite{prst-2016a}.
 These large sample intervals are shown to have good coverage probabilities for nearly all of the distributions listed below in Table~\ref{table1} and $n=$100, 200, 500, 1000  and 5000.
Further the choice of grid size $J=100$ is large enough to obtain this good coverage.
We have developed applications with the R software \citep[R][]{R}, see \cite{prst-2017b} to compute the standard errors and confidence intervals.  For more details see Section \ref{sec:shiny}.

\subsection{Estimating components of $I$}\label{sec:discdecomp}

Given a symmetric $K$-partition with $k$th element $A_k=[p_{k-1},p_k)\cup (1-p_k,1-p_{k-1}]$, we found that for $w_k=2(p_k-p_{k-1})$ the $k$th conditional inequality is determined by (\ref{eqn:Ik}), which is simply
$1-I_k=\bigl\{\int_{2p_{k-1}}^{2p_k}R(u)\,du \bigr\}/w_k.$

We can estimate $I_k$ by first estimating the integral of $R$ over the interval $[2p_{k-1},2p_k]$. To this end define a grid on it by $p_{k_j}=2p_{k-1}+w_k(j - 1/2)/J$ for $j=1,\dots, J.$ (There does not seem to be any benefit in allowing $J$ to
depend on $k$.)  Our estimate of $I_k$ is then
\begin{equation}\label{eqn:Ikhat}
\widehat I_k^{(J)}=\frac{1}{w_k\,J}\sum _{j=1}^J[1 - \widehat{R}(p_{k_j})]~,
\end{equation}
where $\widehat{R}(p_{k_j})$ is an estimate of the quantile ratio $R(p_{k_j}) = Q(p_{k_j}/2)/Q(1-p_{k_j}/2).$
The nominal 100$(1-\alpha)$\% confidence interval for $I_k$ is then
\begin{equation}\label{eqn:ciIk}
\widehat{I}_k^{(J)} \pm z_{1-\alpha/2}\{\widehat{\var}(\widehat{I}_k^{(J)})\}^{1/2}~.
\end{equation}
Details of the formula for the asymptotic variance $\var(\widehat{I}_k^{(J)})$ of $\widehat{I}_k^{(J)}$
and how to estimate it are essentially the same as those for $\widehat{I}^{(J)}$ found in
\citet[Appendix A]{prst-2017b} and so are omitted here.

\marginpar{Table 1 here.}

Simulated coverage probabilities for the interval in \eqref{eqn:ciIk} for varying sample sizes $n$ from 16 different
possible income distributions $F$ are listed in Table~\ref{table1} for the quartile partition.  A total of 1000 trials were conducted for each choice of $n$ and distribution and the nominal coverage was set to 0.95. With the exception of the extreme U-shaped Beta(0.1, 0.1) distribution, excellent coverages are achieved even for the smaller sample size $n=100$ for estimation of both $I_1$ and $I_2$.  Coverage tends to be slightly conservative and is typically closer to nominal for larger sample sizes.  The simulations were repeated with $J=50$ and $J=200$ and similar results were obtained.

For the {\em quintile} partition, simulated coverage probabilities are presented in Table~\ref{table7} in Appendix~\ref{app2}.   These simulations and others not shown here, convince us that the interval estimators of the $I_k$s are reliable for a wide number of income  distributions $F$.

An estimate of $I$ itself can be obtained by applying the decomposition (\ref{eqn:Ihatdecomp}) to the $\hat I_k$s,
but its standard error is not readily obtainable from the standard errors of the components, because although
vector $\widehat {\bI} =(\hat I_1,\dots ,\hat I_K)$ is asymptotically multivariate normal, its limiting covariance matrix $\lim _{n\to\infty }n^{1/2}\cov (\widehat {\bI})$   is not diagonal.

Exact expressions for $\widehat I$ and subcomponents $\widehat I_k$ based on ordered data are also available and
are presented in Appendix~\ref{app3}.

\section{Applications of QRI decompositions}\label{sec:applics}

\subsection{Example 1:\ Australian disposable weekly income}\label{sec:ex1}

Measuring household and personal weekly income is a complicated task carried out by governmental departments, including the Australian Bureau of Statistics (ABS), whose reports are available at \cite{ABSincomedata}.
The gross household income per week is published, but  households differ so much in size
that  the {\em equivalized} disposal weekly income (DWI) is also found. The ABS
defines the DWI as  \lq ... the amount of disposable cash income that a single person household would require to maintain the same standard of living as the household in question, regardless of the size or composition of the latter.\rq\

Table~\ref{table8} in Appendix~\ref{app4} provides ABS grouped data on DWI for selected years, based on representative samples of households
converted to 2014 dollar values. Figure~\ref{fig3} is our depiction of these data with kernel density plots. They are multi-modal distributions and reveal a clear shift to the right of  DWI values over the period 2004--2014 (the solid line with the highest mode is for 2004). We are
interested in tracking inequality over this period.

\marginpar{Figure 3 here.}

To understand how we constructed the plots in Figure~\ref{fig3}, we need to examine Table~\ref{table8} in some detail.
For a partition of 29 classes of dollar incomes listed in the left-most column, and selected financial years, each table entry gives the estimated number of DWIs (in thousands). All amounts have been converted to 2013-2014 dollars using the Consumer Price Index. The financial year in Australia ranges from 1 July of one year to 30 June of the next; for simplicity we hereafter write \lq 2004\rq\  for \lq 2003-2004\rq, and similarly for other years..
For the 2004 data the first entry tells us that there were (approximately) 87,300 zero DWIs, the second entry
that there were 94,100 DWIs between 1 and 49 dollars, and for the last class 371,900 DWIs of 2000 dollars or more.
The last class  has lower bound $x_q=2000$, where $q=1-379.1/ 19606.5=0.981.$

Lacking the individual data, we created an {\em ad hoc}  population to take samples from.  We did this by generating 873 zeros, 941 random uniform values from 1 to 49, 497 uniform values from  55 to 99,
and so on.  For the last category we generated 3,719 random Pareto$(a,\lambda )$ values as follows:\quad first, for a given shape parameter $a>0$, we computed the scale parameter
$\lambda =x_q/\{(1-q)^{-1/a}-1\}=2000/\{(1-0.981)^{-1/a}-1\}$; secondly, we generated 3,719 uniform values $u_i$ from [0.981,1]; and thirdly, we applied the quantile function to these values $Q_{a,\lambda }(u_i)=\lambda\{(1-u_i)^{-1/a}-1\}.$

Standard kernel density plots for the five populations were generated in this way using the default {\tt density} command on R \cite{R}, one for each of the selected years, and these are shown in Figure~\ref{fig3} when the choice of Pareto shape parameter $a=4.$  They are truncated at income 2500, although
their maxima can be much larger, as shown in Table~\ref{table2}.  It is evident that the distributions are moving to the right.  In fact the distribution of 1996 DWI (not shown) is unimodal with mode near 500, while these populations
have two or more modes.  The 2014 density plot is  very similar to Graph 2 of the section \lq Household Income and Wealth Distribution\rq\  \citet{ABSincomedata}.

The ABS also provides the relative standard error (RSE), defined as standard error of  estimate divided by the estimate, for most of their results. For example, in the same source \citet[Table 1.3]{ABSincomedata} from which our Table~\ref{table8} was excerpted, they give the RSE = 11.8\%  for  the first table entry 87.3 (thousands of  persons with zero DWI in 2004); that is, the standard error is nearly 10.3 thousand.  For later years where the sample number of households was much higher, such as in 2014, the RSE's are in the range 5--8 \%.  The main point is that even with
the best of survey methods, the resulting summary data listed in the ABS Table~1.3 only approximately describes
the exact populations of DWI. Our five populations, truncated at \$2500 to fit in Figure~\ref{fig3}, are also approximations.

The percentiles  of our five populations are listed in Table~\ref{table2}. They are in good agreement
with the percentile estimates of \citet[Table 1.1]{ABSincomedata}. For example, for the year 2004
they obtain P10=324, P20=395, P50=657, P80=1,008 and P90=1,255. And for 2014 they obtain P10=415, P20=511,
P50=844, P80=1308 and P90=1688.

The maximum values in the second-last column
of Table~\ref{table2} could have been quite different, (and much larger for $a$ near 1). However, extremely large incomes (outliers) do not affect the quantile ratio index estimator $\widehat I$ very much.
Note that $\widehat I$, also listed in Table~\ref{table2}, does not appear to be changing over the years 2004--2014,
indicating a stable level of inequality for this period.  Changing the Pareto tail shape  $a$ from 4 to 1 greatly increases the maximum values in this table, but has no effect on the other quantiles and little effect on $\hat I.$

\marginpar{Table 2 here.}

\subsubsection*{Quartile partition estimates for DWI.}\label{sec:DWIquartile}
Next consider the \lq quartile partition\rq\ with members $A_1=[0,0.25]\cup [0.75,1]$ and $A_2=[0.25,0.75].$
Incomes in $A_2$ can be considered as belonging to the \lq middle class\rq.\   Here $\widehat I_k$ is defined by
(\ref{eqn:Ikhat}) for $k=1,2.$  These estimates, as well as those for $I$ were based on samples of size $n=10,000$
 for each of the five populations generated with $a=4$ and are listed in Table~\ref{table3}.

The bottom row shows the estimates of $I$ are only increasing slightly over time. We have already commented upon this stability earlier for the right-most column of Table~\ref{table2}. However, a
 two-sided level 0.05 test for a  difference between the 2004 and 2014 values of $\widehat I$ is just  significant.
 We can now find the source of this increasing inequality.

The middle row of Table~\ref{table3} shows that inequality in the middle class, measured
by $\widehat I_2$, has not significantly changed over this period.  But the inequality for partition $A_1$, containing the lower quartile and upper quartiles, has increased from 0.721 to 0.736, which is significant at the 0.05 level .  It is interesting that the standard errors of $\widehat I_1$ and $\widehat I$
are roughly the same, while those of $\widehat I_2$ for the middle class are about 1/4 larger.
Similar results were obtained for samples based on $n=4000$ observations, but for $n=1000$ changes in
inequality over this time period were not quite statistically significant at the 0.05 level.

\subsection{Example 2:\ Australian wealth data}\label{sec:ex2}
 In its explanatory notes of Manual 6523.0, the ABS defines household wealth by \lq  Net worth, often
 is the value of a household's assets less the value of its liabilities.\rq\  and then
goes on to explain what it means by assets and liabilities. Table~\ref{table9}  is obtained from
\citet[Table 2.3]{ABSincomedata}.  Because of the large numbers involved, we have written the dollar classes in
thousands of dollars.  For example, the second entry in the column labelled 2004 is  1098.9, which
means that there were an estimated
1,098,900  households in that year whose net wealth was between 0 and \$50,000 in 2014.
We used the methods of Example~1 to generate a population that reflects the information in Table~\ref{table9},
modulo the shape parameter for a Pareto tail for the last unbounded dollar class.
Density plots for the five populations are shown in Figure~\ref{fig4}. They are truncated at 800 for this plot because the graphs are visually almost indistinguishable for larger values. The graph for 2004 is unimodal, while for subsequent years a clear shift to bimodality is apparent.

 Empirical percentiles
for these data in Table~\ref{table4} reveal that while the lower percentiles are not changing over the
decade, the median P50 and larger percentiles appear to be steadily increasing. The estimated inequality of wealth
$\widehat I$ appears to be increasing only slightly over the decade. Nevertheless $\widehat I$ values near 0.7 are certainly much
higher than those for disposable income, which was near 0.5, see the last column of Table~\ref{table2}.
We now examine the possible change in inequality over the Australian population of households and certain sub-populations of these wealth data.

\marginpar{Table 4 here.}

\marginpar{Figure 3 here.}

\subsubsection{Quartile partition estimates for NHW.}\label{sec:NHWquartile}
Next consider the quartile partition with members $A_1=[0,0.25]\cup [0.75,1]$ and $A_2=[0.25,0.75].$
Incomes in the population image of
$A_2$ can be considered as belonging to the \lq middle class\rq.\   Here $\widehat I_k$ is defined by
(\ref{eqn:Ikhat}) for $k=1,2.$  These estimates are found for each of the five populations generated (as was done for the DWI data) starting with the grouped data listed in Table~\ref{table9}. They are based on 10,000 observations
from each of the respective populations.

A level 0.05 test for a  difference between the 2004 and 2014 values of $\widehat I$, namely $|0.726-0.714|=0.012$
would reject for $n=10,000$ because then the standard error of the difference between them is $\se =\sqrt {(0.33^2+0.31^2)/n}\, =0.004$.  A similar test for significant change in inequality for the middle class over the same range of time is just significant at  the 0.05 level, because the difference is 0.018 and the standard error of the difference is 0.008.  The standard errors for the estimates of inequality in the outer quartiles $\widehat {I} _1$
are much smaller, so while the difference between the 2014 and 2004 results is only 0.955-0.949 =0.006 but its standard error is also much smaller at 0.002, leading to a statistically significant result. Thus most of the change in wealth  inequality over this period is due to the change in the lower
and upper quartiles. Note that such tests are correlated.

\marginpar{Table 5 here.}

\subsubsection*{Decile partition estimates for NHW.}\label{sec:NHWdecile}
Next we look at a finer partition, the decile partition, to further pinpoint where the wealth inequality is changing most. Estimates and their standard errors are given in Table~\ref{table6}. Note that the standard errors
of the estimates can vary from 0.0003 to 0.0069.

Using (\ref{eqn:wtedmean}) the results in the second column
show that for 2004 the overall inequality can be broken down into
$ \widehat I=0.707=0.198+0.189+0.162+0.113+0.045$,
so the first three partition members (outer six deciles) contribute 0.549/0.707 or almost 80\% to the overall QRI
and the fifth partition (central two deciles) only 0.045/0.707, or 6\%.

Comparing the estimates of $I_5$ in 2004 and 2014 for the central partition shows a less than one standard error of
increase. However, for every other partition member the QRI has increased by more
than two standard errors over this time period. We can conclude that wealth inequality is becoming  more unequal except within  the central 1/5 of the population.

\marginpar{Table 6 here.}

\section{Shiny applications}\label{sec:shiny}

For convenience, we have created two Shiny \citep{shiny} applications that some readers may find useful.  These applications can be found at
\begin{center}
\url{https://lukeprendergast.shinyapps.io/Decomp/}
\end{center}
and
\begin{center}
\url{https://lukeprendergast.shinyapps.io/QRIestimation/}.
\end{center}
The first calculates the QRI for several distributions considered within this manuscript.  This application also calculates the quartile and quintile decompositions for the QRI if requested.  The second allows the user to upload a csv data file for which to estimate the QRI and its quartile and quintile decompositions.  The standard errors and large sample confidence interval estimators for the QRI and its decompositions are included in the tabulated output.  We will continue to improve these applications and are grateful for any feedback.

\section{Summary}\label{sec:conclusion}

We have learned that for Australian data, inequality of DWI as measured by the QRI is almost steady at about $\widehat I=0.5$ over the past 11 years, although for a large enough sample it has increased by a statistically significant
amount.  Samples of size 1000 would not detect such an increase. Further, by examining the QRI estimates for the quartile partition, we found that  all the change in inequality of incomes over the time
period 2004 to 2014 can be attributed to the outer classes, with incomes inequality in the central class remaining
stagnant. The inequality in wealth NHW for Australian households over this same period was much higher $\widehat I \approx 0.7$ and very significantly increases from 2004 to 2014.   Moreover, hardly any of this increase is due to the middle two deciles.

These examples illustrate the simple utility of measuring inequality with the QRI.  Not only can we
find reliable confidence intervals for the QRI of a population using relatively small samples in the hundreds,
we can also find them for symmetric partitions of quantiles, and use them to discern those which contribute most
to the population QRI and how much them contribute.  Finally we can use these results to detect changes over time.
Given  income or wealth data from various countries, it would be straightforward, using the accompanying
programs in Section~\ref{sec:shiny}, to compute the QRI for each of them and/or desired symmetric sub-populations. One could also study
income data for those in the top 10\%, or any other region of interest.

\bibliography{prstbib}
\bibliographystyle{authordate4}


Starting with  $R_\sigma (p)=\exp \{2\sigma z_{p/2}\}$ and making  the change of variable $y=2\sigma z_{p/2}$
\[\int _0^rR_\sigma (p)\,dp =\frac{1}{\sigma }\int _{-\infty }^{2\sigma z_{r/2}} e^y\,\varphi\left (\frac{y}{2\sigma }\right )\,dy ~,\]
where $\varphi(z)=e^{-z^2/2}/\sqrt{2\pi}$ is the standard normal density.
Completing the square within  the exponential of the integrand $\exp \{y-(y/2\sigma )^2/2\}$ leads to
\[  \frac {8\sigma ^2y-y^2}{8\sigma ^2}=\frac {16\sigma ^4-(y-4\sigma ^2)^2}{8\sigma ^2}~, \text { so }\]
\[\int _0^rR_\sigma (p)\,dp =2\exp \{2\sigma ^2\}\,\int _{-\infty }^{2\sigma z_{r/2}}  \frac {1}{2\sigma}\varphi \left (\frac {y-4\sigma ^2}{2\sigma }\right )\,dy =  2\exp \{2\sigma ^2\}\,\Phi (\Phi^{-1}(r/2)-2\sigma )~.\]

\vspace{2cm}

\subsection{Appendix 2. Coverage probabilities for the quintile partition}\label{app2}

\marginpar{Table 7 here.}

\subsection{Appendix 3. Exact ecomposition formula for $\hat I$}\label{app3}

Taking the weighted average of the $\widehat{I}^{(J)}_k$s defined in \eqref{eqn:Ikhat} does not guarantee that it will be exactly equal to the estimator of $I$  found by $\widehat I^{(J)}=\{\sum _{j=1}^J[1 - \widehat{R}(p_{j})\}/J$.  However, \cite{prst-2017b} showed the estimates are stable for moderate to large choices of $J$ so that the resulting weighted average of the $\widehat{I}^{(J)}_k$s is expected to be very close to $\widehat{I}^{(J)}$.  However, if $n_0=np_0=0,$ $n_1=np_1$, $n_2=np_2 \dots , n_{K-1}=np_{K-1}$ and $n_K= n/2$ are all distinct integers, then it is possible to define estimates of the $\widehat{I}_k$s such that their weighted average is equal to a simple estimator of $I$.

Given ordered incomes $0 \leq x_{1}\leq  x_{2}\leq \dots \leq x_{n}$, where  $n\geq 2$ and the frequency of 0's is less than $0.5,$ let $k= \lfloor n/2\rfloor .$  It is shown in \citet[Equation 3]{prst-2017b}
 that an exact estimate of $I$ is given by
\begin{equation}\label{eqn:exactI}
   I_n=I(F_n)= \frac{2}{n}\,\sum _{j=1}^{k}\left (1-\frac {x_{j}}{x_{n-j+1}}\right )~.
\end{equation}

 Given  a symmetric $K$-partition $\{A_1,\dots ,A_K\}$ of the unit interval determined by $0=p_0< p_1<\dots <p_K=1/2,$ we want to decompose $\widehat I$ into a weighted average of individual estimates $\widehat I_k$ of $I_k$.
To this end assume $n_0,n_2,\ldots$ as above and assume that they are distinct integers.
Define the disjoint sets $B_k=\{n_{k-1}+1,n_{k-1}+2,\dots ,n_k\},$ for $k=1,\dots,K.$ The length of $B_k$ is $m_k=n_k-n_{k-1}$ and $\sum _{k=1}^Km_k=n_K=n/2.$
We  estimate $I_k$ by
\begin{equation*}
  \widehat I_k = \frac{1}{m_k}\,\sum _{j\in B_k}\left (1-\frac {x_{j}}{x_{n-j+1}}\right )~.
\end{equation*}
Then a simple estimator of $\widehat{I}$ based on the order statistics can be written
\begin{eqnarray}\label{eqn:Ihatdecomp}\nonumber
 \widehat I &=& \frac{2}{n}\,\sum _{j=1}^{n/2}\left (1-\frac {x_{j}}{x_{n-j+1}}\right )\\ \nonumber
       &=& \frac{2}{n}\, \sum _{k=1}^K \sum _{j\in B_k}\left (1-\frac {x_{j}}{x_{n-j+1}}\right )\\
        &=& \frac{2}{n}\,\sum _{k=1}^K  m_k\,\widehat I_k  ~,
\end{eqnarray}
where $m_k= n_k-n_{k-1}.$ Note that the sum of the weights $\sum _k 2m_k/n= 1.$

\subsection{Appendix 4. Australian Bureau of Statistics Income Data}\label{app4}

\marginpar{Table 8 here.}

\subsection{Appendix 5. Australian Bureau of Statistics Wealth Data}

\marginpar{Table 9 here.}

\clearpage
\newpage

\begin{table}[ht]\centering
\caption{Simulated coverage probabilities for the quartile partition with sample sizes $n=100,500$ and 1000 for various choices of standard income distributions $F$, also studied in \citet[Tables 1 and 3]{prst-2017b}.
 A total of 1000 trials were conducted.}\label{table1}
\begin{tabular}{clcccccc}
  \hline
    && \multicolumn{2}{c}{$n=100$} & \multicolumn{2}{c}{$n=500$} & \multicolumn{2}{c|}{$n=1000$} \\
  \# & $F$ & $I_1$ & $I_2$  & $I_1$  & $I_2$  & $I_1$  & $I_2$  \\
  \cline{1-2}  \cline{3-4} \cline{5-6} \cline{7-8}
  1 & Lognormal      & 0.966 & 0.960 & 0.970 & 0.968 & 0.955 & 0.957 \\
  2 & Beta(0.1,0.1)  & 1.000 & 0.918 & 1.000 & 0.976 & 1.000 & 0.965 \\
  3 & Beta(0.5,0.5)  & 0.930 & 0.926 & 0.940 & 0.941 & 0.952 & 0.927 \\
  4 & Beta(1,1)      & 0.943 & 0.935 & 0.953 & 0.946 & 0.957 & 0.952 \\
  5 & Beta(10,10)    & 0.969 & 0.972 & 0.976 & 0.960 & 0.972 & 0.961 \\
  6 & $\chi^2_1$     & 0.957 & 0.973 & 0.961 & 0.964 & 0.946 & 0.940 \\
  7 & $\chi^2_4$     & 0.966 & 0.959 & 0.963 & 0.955 & 0.960 & 0.958 \\
  8 & $\chi^2_{25}$  & 0.974 & 0.965 & 0.973 & 0.955 & 0.959 & 0.945 \\
  9 & Pareto(1)      & 0.966 & 0.989 & 0.966 & 0.963 & 0.959 & 0.962 \\
  10 & Pareto(2)     & 0.965 & 0.976 & 0.962 & 0.960 & 0.954 & 0.958 \\
  11 & Pareto(100)   & 0.952 & 0.965 & 0.955 & 0.969 & 0.956 & 0.948 \\
  12 & Exp(1)        & 0.956 & 0.963 & 0.945 & 0.960 & 0.949 & 0.962 \\
  13 & Weibull(0.5)  & 0.968 & 0.993 & 0.967 & 0.969 & 0.962 & 0.969 \\
  14 & Weibull(2)    & 0.959 & 0.961 & 0.966 & 0.953 & 0.960 & 0.957 \\
  15 & Weibull(10)   & 0.980 & 0.983 & 0.985 & 0.955 & 0.987 & 0.960 \\
  16 & LN-Frechet    & 0.983 & 0.970 & 0.973 & 0.968 & 0.968 & 0.959 \\
   \hline
\end{tabular}
\end{table}

\begin{table}[h!]\centering
\caption{\label{table2} {Percentiles for five distributions of DWI depicted in Figure~\ref{fig3}, and
values of $\widehat I$ rounded to two places. The minimum value for each distribution was 0.}}
\begin{tabular}{ccccccccccccc}
 \hline
      & P05 & P10 & P20 & P25 & P50 & P75  &  P80 &  P90 &  P95 & $\max$ & $\widehat I$ \\
\hline
2004  & 269 & 320 & 394 & 433 & 658 &  928 & 1008 & 1255 & 1521 & 33520  & 0.51\\
2006  & 292 & 340 & 426 & 472 & 707 & 1003 & 1096 & 1383 & 1714 & 39441  & 0.51\\
2010  & 309 & 374 & 470 & 526 & 793 & 1163 & 1273 & 1615 & 2024 & 39666  & 0.52\\
2012  & 317 & 396 & 497 & 552 & 831 & 1188 & 1298 & 1642 & 1989 & 34226  & 0.52\\
2014  & 321 & 411 & 509 & 558 & 843 & 1196 & 1309 & 1688 & 2179 & 32970  & 0.52\\
\hline
\end{tabular}
\end{table}

\begin{table}[t!]\centering
\caption{\label{table3} {Estimates of $I$ and $I_k$ for the quartile partition of the five distributions of
Figure~\ref{fig4}, based on samples of size $n=10,000$. In parentheses are the values of $\sqrt n\ \widehat {\se } [\widehat I_k].$  }}
\begin{tabular}{cccccc}
\hline
                &  2004        &     2006          & 2010         & 2012         &  2014     \\
\hline
 $\widehat I_1$ &  0.721(0.27) &  0.722(0.27)   &  0.742(0.27) &  0.733(0.28) &  0.736(0.28)  \\
 $\widehat I_2$ &  0.285(0.34) &  0.289(0.33)   &  0.297(0.34) &  0.285(0.34) &  0.287(0.34)  \\
 $\widehat I$   &  0.503(0.27) &  0.506(0.26)   &  0.520(0.27) &  0.509(0.27) &  0.512(0.27)   \\
\hline
  \end{tabular}
\end{table}

\begin{table}[b!]\centering
\caption{\label{table4} {Percentiles for five distributions of NHW depicted in Figure~\ref{fig4}, and
values of $\widehat I$. The minimum value for each distribution was 0.}}
\begin{tabular}{ccccccccccccc}
\hline
      & P05 & P10 & P20 & P25 & P50 &  P75 &  P80 &  P90 &  P95 &   max  &  $\widehat I$ \\
\hline
2004  &  15 &  32 & 86  & 132 & 388 & 747  &  867 & 1319 & 1984 & 43203  &  0.71     \\
2006  &  15 &  34 & 89  & 137 & 424 & 787  &  942 & 1477 & 2182 & 42635  &  0.71     \\
2010  &  16 &  35 & 97  & 147 & 474 & 909  & 1070 & 1670 & 2567 & 62912  &  0.72     \\
2012  &  15 &  34 & 93  & 140 & 458 & 913  & 1085 & 1697 & 2526 & 40650  &  0.73     \\
2014  &  16 &  36 & 95  & 143 & 463 & 961  & 1129 & 1770 & 2717 & 41328  &  0.73     \\
\hline
\end{tabular}
\end{table}

\begin{table}[h!]
\centering
\caption{\label{table5} {Estimates of $I_k$ for the quartile partition of the five distributions of
Figure~\ref{fig4} based on 10,000 observations selected at random from each of them.
 In parentheses are the values of $\sqrt n\ \widehat {\se }[\widehat I_k]$.}}
\vspace{.1cm}
\begin{tabular}{cccccc}
\hline
            &   2004        &    2006       &    2010      &    2012       &     2014      \\
 $\widehat I_1$ &   0.949(0.16) &   0.948(0.16)  &  0.952(0.16) &   0.952(0.15) &   0.955(0.14) \\
 $\widehat I_2$ &   0.480(0.54) &   0.466(0.55)  &  0.473(0.56) &   0.500(0.53) &   0.498(0.53) \\
 $\widehat I$   &   0.714(0.33) &   0.707(0.33)  &  0.712(0.34) &   0.726(0.32) &   0.726(0.31) \\
\hline
 \end{tabular}
\end{table}

\begin{table}[t!]\centering
\caption{\label{table6} Estimates of $I_k$ for the decile partitioning of the five distributions deicted in
Figure~\ref{fig4}. As in Table~\ref{table5}, estimates are  based on 10,000 observations selected at random from each of them, and values in parentheses are $100\times \widehat {\se }[\widehat I_k]$.}
\begin{tabular}{cccccc}
\hline
                &  2004       &     2006       &     2010     &     2012      &  2014         \\
\hline
 $\widehat I_1$ & 0.991(0.04) &   0.991(0.04)  &  0.992(0.03) &   0.992(0.03) &   0.993(0.03) \\
 $\widehat I_2$ & 0.944(0.19) &   0.945(0.19)  &  0.949(0.18) &   0.949(0.18) &   0.951(0.16) \\
 $\widehat I_3$ & 0.812(0.51) &   0.811(0.52)  &  0.827(0.48) &   0.832(0.44) &   0.842(0.42) \\
 $\widehat I_4$ & 0.565(0.69) &   0.557(0.69)  &  0.584(0.68) &   0.606(0.69) &   0.618(0.66) \\
 $\widehat I_5$ & 0.223(0.51) &   0.218(0.49)  &  0.232(0.52) &   0.235(0.54) &   0.229(0.55) \\
 $\widehat I$   & 0.707(0.33) &   0.705(0.33)  &  0.717(0.32) &   0.723(0.32) &   0.729(0.31) \\
\hline
  \end{tabular}
\end{table}

\begin{table}[ht]
\centering
 \caption{Simulated coverage probabilities for the quintile partition with sample sizes $n=100,500$ and 1000.  A total of 1000 trials were conducted.}\label{table7}
\small
\begin{tabular}{clccccccccc}
  \hline
      && \multicolumn{3}{c}{$n=100$} & \multicolumn{3}{c}{$n=500$} & \multicolumn{3}{c|}{$n=1000$} \\
\# & $F$ & $I_1$ & $I_2$ & $I_3$ & $I_1$ & $I_2$ & $I_3$ & $I_1$ & $I_2$ & $I_3$ \\
  \cline{1-2}  \cline{3-5} \cline{6-8} \cline{9-11}
  1  & Lognormal      & 0.976 & 0.965 & 0.964 & 0.966 & 0.958 & 0.959 & 0.968 & 0.969 & 0.943 \\
  2  & Beta(0.1,0.1)  & 0.999 & 0.991 & 0.799 & 1.000 & 1.000 & 0.882 & 1.000 & 0.999 & 0.899 \\
  3  & Beta(0.5,0.5)  & 0.926 & 0.934 & 0.935 & 0.937 & 0.937 & 0.941 & 0.944 & 0.933 & 0.925 \\
  4  & Beta(1,1)      & 0.946 & 0.947 & 0.951 & 0.957 & 0.954 & 0.953 & 0.948 & 0.938 & 0.947 \\
  5  & Beta(10,10)    & 0.955 & 0.968 & 0.961 & 0.984 & 0.960 & 0.948 & 0.975 & 0.970 & 0.961 \\
  6  & $\chi^2_1$     & 0.955 & 0.967 & 0.979 & 0.947 & 0.944 & 0.958 & 0.962 & 0.963 & 0.955 \\
  7  & $\chi^2_4$     & 0.973 & 0.962 & 0.959 & 0.953 & 0.945 & 0.943 & 0.971 & 0.953 & 0.952 \\
  8  & $\chi^2_{25}$  & 0.957 & 0.974 & 0.969 & 0.978 & 0.954 & 0.955 & 0.971 & 0.960 & 0.959 \\
  9  & Pareto(1)      & 0.968 & 0.982 & 0.987 & 0.967 & 0.962 & 0.959 & 0.962 & 0.958 & 0.966 \\
  10 & Pareto(2)      & 0.949 & 0.966 & 0.972 & 0.950 & 0.948 & 0.955 & 0.962 & 0.968 & 0.955 \\
  11 & Pareto(100)    & 0.955 & 0.968 & 0.958 & 0.958 & 0.962 & 0.949 & 0.948 & 0.954 & 0.950 \\
  12 & Exp(1)         & 0.944 & 0.965 & 0.971 & 0.960 & 0.952 & 0.947 & 0.942 & 0.948 & 0.952 \\
  13 & Weibull(0.5)   & 0.969 & 0.989 & 0.996 & 0.964 & 0.966 & 0.973 & 0.956 & 0.969 & 0.958 \\
  14 & Weibull(2)     & 0.966 & 0.960 & 0.962 & 0.971 & 0.940 & 0.940 & 0.956 & 0.958 & 0.954 \\
  15 & Weibull(10)    & 0.973 & 0.990 & 0.961 & 0.985 & 0.971 & 0.954 & 0.979 & 0.957 & 0.955 \\
  16 & LN-Frechet     & 0.985 & 0.979 & 0.980 & 0.977 & 0.974 & 0.967 & 0.971 & 0.964 & 0.953 \\
   \hline
\end{tabular}
\end{table}

\renewcommand{\baselinestretch}{1}

\begin{table}[h]\begin{center}
\caption{\label{table8}Equivalized disposable weekly income (DWI)  in Australian dollars, adjusted for inflation
to 2013-2014 dollars, for selected financial years. The tabled entries represent thousands of persons. Source: \citet[Table 1.3]{ABSincomedata}, downloaded 27 July, 2017. For further analysis of this source, see \cite{wilkins-2015}.}
\begin{tabular}{lrrrrr}
 \hline
                        & 2003--2004 &2005--2006 &   2008--2010 & 2011--2012 &  2013--2014 \\
  \hline
 $[0,0]$ $^a$           &   87.3     &   73.7 &   89.0 &   87.4 &   86.4    \\
 $[1,49]$               &   94.1     &   90.1 &   95.8 &   81.6 &   95.3    \\
 $[50,99]$              &   49.7     &   63.1 &   61.3 &   85.3 &   78.9    \\
 $[100,149]$            &   94.0     &   66.2 &   84.0 &   92.3 &   47.6    \\
 $[150,199]$            &  129.9     &  108.6 &  125.1 &  107.3 &  134.9    \\
 $[200,249]$            &  273.7     &  219.6 &  164.7 &  185.6 &  151.1    \\
 $[250,299]$            &  657.6     &  443.7 &  351.5 &  335.0 &  373.4    \\
 $[300,349]$            & 1385.5     & 1152.0 &  596.3 &  373.9 &  397.9    \\
 $[350,399]$            & 1301.8     & 1187.5 & 1195.8 &  913.3 &  636.7    \\
 $[400,449]$            & 1231.7     & 1111.8 & 1172.4 & 1184.1 & 1135.2    \\
 $[450,499]$            & 1093.7     & 1052.3 &  933.4 & 1044.7 & 1175.2    \\
 $[500,549]$            & 1043.0     & 1097.4 &  991.3 & 1019.7 & 1171.7    \\
 $[550,599]$            & 1092.2     & 1057.0 & 1009.7 &  980.8 & 1093.0    \\
 $[600,649]$            & 1087.5     & 1016.2 & 1046.4 &  926.3 &  956.6    \\
 $[650,699]$            & 1083.5     & 1066.9 &  987.0 & 1021.9 &  972.7    \\
 $[700,749]$            & 1092.8     & 1023.3 &  996.9 &  999.2 &  938.9    \\
 $[750,799]$            &  959.9     &  834.1 & 1037.1 & 1038.1 & 1009.6    \\
 $[800,849]$            &  878.1     &  940.4 &  829.3 &  989.4 & 1013.4    \\
 $[850,899]$            &  718.3     &  828.5 &  806.5 &  959.7 & 1099.5    \\
 $[900,949]$            &  612.2     &  746.6 &  793.0 &  896.4 &  826.2    \\
 $[950,999]$            &  631.8     &  731.9 &  757.8 &  714.9 &  885.6    \\
 $[1000,1049]$          &  506.8     &  547.5 &  630.3 &  690.1 &  692.6    \\
 $[1050,1099]$          &  492.3     &  515.3 &  730.8 &  803.1 &  695.8    \\
 $[1100,1199]$          &  750.3     &  933.9 & 1118.5 & 1245.7 & 1379.5    \\
 $[1200,1299]$          &  529.4     &  674.2 &  906.1 &  985.3 & 1027.2    \\
 $[1300,1499]$          &  706.4     &  863.9 & 1400.8 & 1499.2 & 1447.8    \\
 $[1500,1699]$          &  387.9     &  469.6 &  889.7 &  995.4 &  938.5    \\
 $[1700,1999]$          &  263.2     &  427.0 &  682.8 &  850.2 &  862.3    \\
 $[2000, +\infty)$ $^b$ &  371.9     &  588.4 & 1106.3 & 1082.9 & 1355.6     \\
\hline
Total                   &  19,606.5  & 19,930.7 & 21,589.6 & 22,188.8 & 22,679.1 \\
\hline
\end{tabular}
\end{center}
\noindent ${a.}$ Some DWIs are negative, but these values have been rounded up to zero.

\noindent ${b.}$ An upper bound on DWIs greater than \$2000 is not reported.

\noindent ABS caveat: \lq\lq ..estimates presented for 2007-–08  onwards are not directly comparable with estimates for previous cycles due to the improvements made to measuring income introduced in the 2007–-08 cycle. Estimates for 2003–-04 and 2005-–06 have been recompiled to reflect the new measures of income, however not all components introduced in 2007-–08 are available for earlier cycles.\rq \rq\ 													
\end{table}

\begin{table}[h]\begin{center}
\small
\caption{\label{table9}Net Household Wealth (NHW) in thousands of Australian dollars, adjusted for inflation to 2013-2014 dollars, for all available financial years. The tabled entries represent thousands of households. Source: \citet[Table 2.3]{ABSincomedata}, downloaded 27 July, 2017.}
\begin{tabular}{lrrrrr}
\hline
                         &  2003--2004 & 2005--2006 & 2009--2010  & 2011--2012  & 2013--2014 \\
  \hline
 $(-\infty,0)$ $^a$      &    56.6     &    75.6    &    77.3     &   113.7    &    93.8    \\
 $[   0,  49]$           &  1098.9     &  1044.6    &  1058.2     &  1075.2    &  1052.7    \\
 $[  50,  99]$           &   547.0     &   583.4    &   577.4     &   617.1    &   667.3    \\
 $[ 100, 149]$           &   364.3     &   374.1    &   408.5     &   441.9    &   433.3    \\
 $[ 150, 199]$           &   365.4     &   308.0    &   311.5     &   368.5    &   334.1    \\
 $[ 200, 249]$           &   372.7     &   305.4    &   294.6     &   337.7    &   360.1    \\
 $[ 250, 299]$           &   393.3     &   354.3    &   309.1     &   319.8    &   363.1    \\
 $[ 300, 349]$           &   372.8     &   356.5    &   348.6     &   306.4    &   329.5    \\
 $[ 350, 399]$           &   397.5     &   397.0    &   331.0     &   335.9    &   343.3    \\
 $[ 400, 449]$           &   353.5     &   351.6    &   332.1     &   360.7    &   329.3    \\
 $[ 450, 499]$           &   335.5     &   361.8    &   348.2     &   333.0    &   342.4    \\
 $[ 500, 599]$           &   574.9     &   601.4    &   621.8     &   554.7    &   570.4    \\
 $[ 600, 699]$           &   402.6     &   492.9    &   508.6     &   499.4    &   451.0    \\
 $[ 700, 799]$           &   365.9     &   400.8    &   425.8     &   410.4    &   420.2    \\
 $[ 800, 899]$           &   295.6     &   252.9    &   344.0     &   377.3    &   325.8    \\
 $[ 900, 999]$           &   211.1     &   220.5    &   283.1     &   267.5    &   298.0    \\
 $[1000,1099]$           &   179.5     &   189.5    &   235.5     &   234.0    &   261.9    \\
 $[1100,1199]$           &   147.3     &   161.9    &   185.0     &   203.9    &   202.4    \\
 $[1200,1399]$           &   233.7     &   241.9    &   314.7     &   323.6    &   346.4    \\
 $[1400,1599]$           &   138.3     &   196.4    &   213.1     &   228.6    &   242.9    \\
 $[1600,1799]$           &    92.8     &   122.6    &   161.8     &   158.6    &   171.5    \\
 $[1800,1999]$           &    69.9     &    83.7    &   118.9     &   153.3    &   149.0    \\
 $[2000,2199]$           &    64.9     &    77.4    &    75.6     &    94.7    &   100.8    \\
 $[2200,2399]$           &    50.9     &    53.8    &    72.3     &    68.0    &    78.8    \\
 $[2400,2599]$           &    30.1     &    44.4    &    62.3     &    55.6    &    62.0    \\
 $[2600,2999]$           &    55.8     &    73.6    &    98.4     &    87.6    &    91.4    \\
 $[3000,3999]$           &    83.6     &    90.1    &   111.1     &   126.3    &   149.2    \\
 $[4000,4999]$           &    19.4     &    41.2    &    61.8     &    68.7    &    70.5    \\
 $[5000,6999]$           &    34.8     &    36.3    &    51.7     &    55.3    &    63.9     \\
 $[7000,9999]$           &    11.7     &    14.5    &    25.4     &    33.1    &    31.8     \\
 $[10000,+\infty )$ $^b$ &    15.5     &    18.3    &    31.4     &    19.9    &    29.5     \\
\hline
Total                    &  7,735.8    & 7,926.4   & 8,398.8     & 8,630.4    &  8,766.3  \\
\hline
\end{tabular}
\end{center}
\noindent ${a.}$ The unknown NHWs less than 0 will be assigned 0 in our analysis.

\noindent ${b.}$ An upper bound on NHWs greater than \$10,000,000.00 is not reported.
\end{table}

\clearpage
\newpage

 \begin{figure}[t!]
\begin{center}
\caption{Plots of the quantile ratio $R(p)=Q(p/2)/Q(1-p/2)$ for the standard lognormal distribution $F$ and, for the
quintile partition, $R_k(p)$ defined by (\ref{eqn:Rk}) for $k=1,2,3$. The inequality measure $I_k$ for each $R_k$ is the shaded area above its graph and below the horizontal line at one.  \label{fig1}}
\end{center}
\end{figure}

\begin{figure}[t!]
\begin{center}
\caption{Plots in solid lines of the graphs of $I$ versus the shape parameters for four
standard families. For the quartile partition, $I=(I_1+I_2)/2$, and the dashed lines show the respective
graphs of $I_1/2$ defined by (\ref{eqn:Ik})  and the dotted lines those of $I_2/2$.
\label{fig2}}
\end{center}
\end{figure}

\begin{figure}[b]
\caption{Kernel density estimates for the five populations generated in Section~\ref{sec:ex1}, truncated to
[0,2500].
The small positive mass on 0 is smoothed out by these density estimates. \label{fig3}}
\end{figure}

\begin{figure}[t!]
\caption{Kernel density estimates for the five populations of wealth data of Table~\ref{table9}, generated as in Section~\ref{sec:ex1}, and truncated to [0,800]. The solid line with the single, highest mode is the graph for 2004; it crosses the graph for 2014, also in solid line, which depicts a much more dispersed population of incomes that is bimodal.\label{fig4}}
\end{figure}

\clearpage
\newpage

{\bf Figure 1}

\begin{figure}
\begin{center}
\includegraphics[scale=.6]{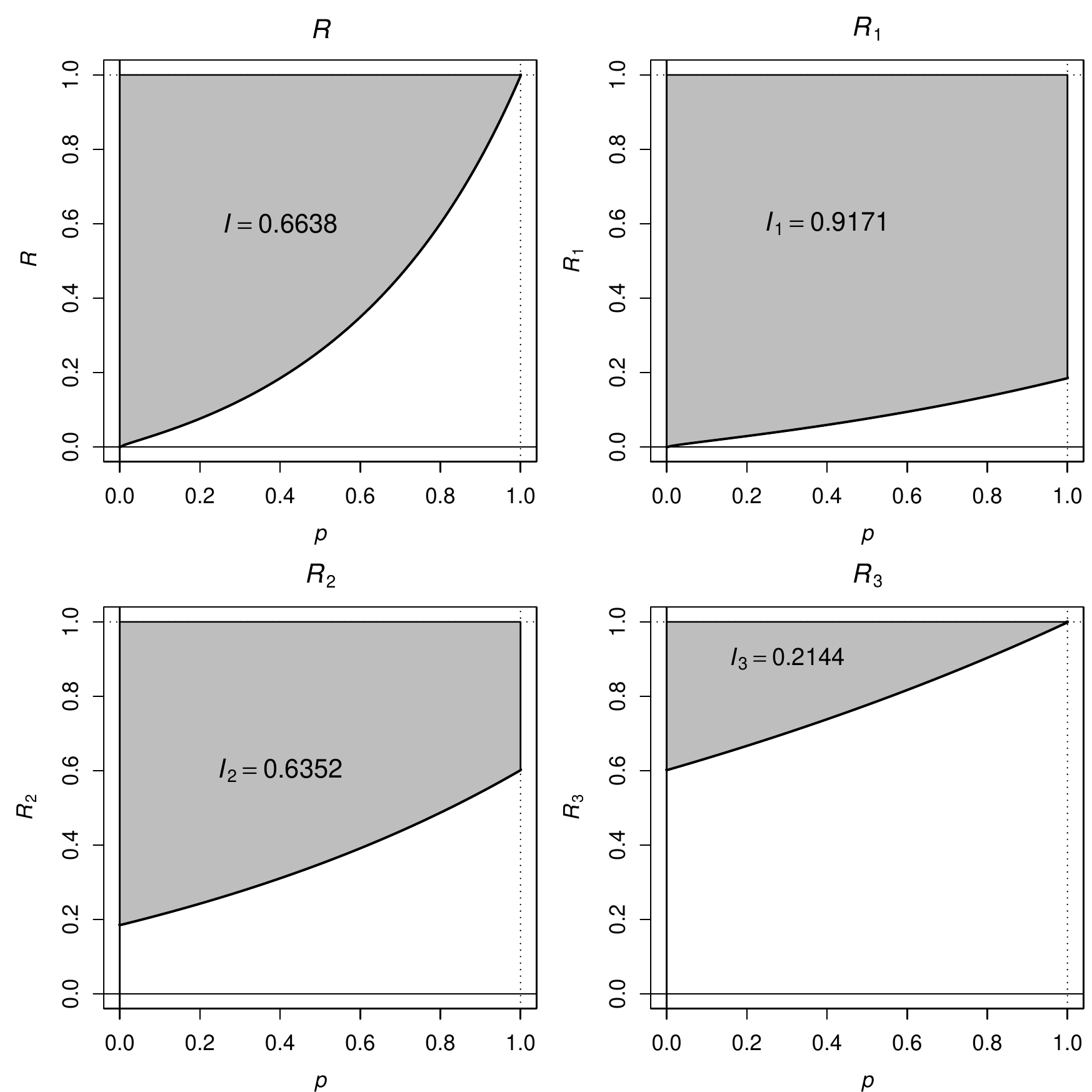}
\end{center}
\end{figure}

\clearpage
\newpage

{\bf Figure 2}

\begin{figure}
\begin{center}
\includegraphics[scale=.6]{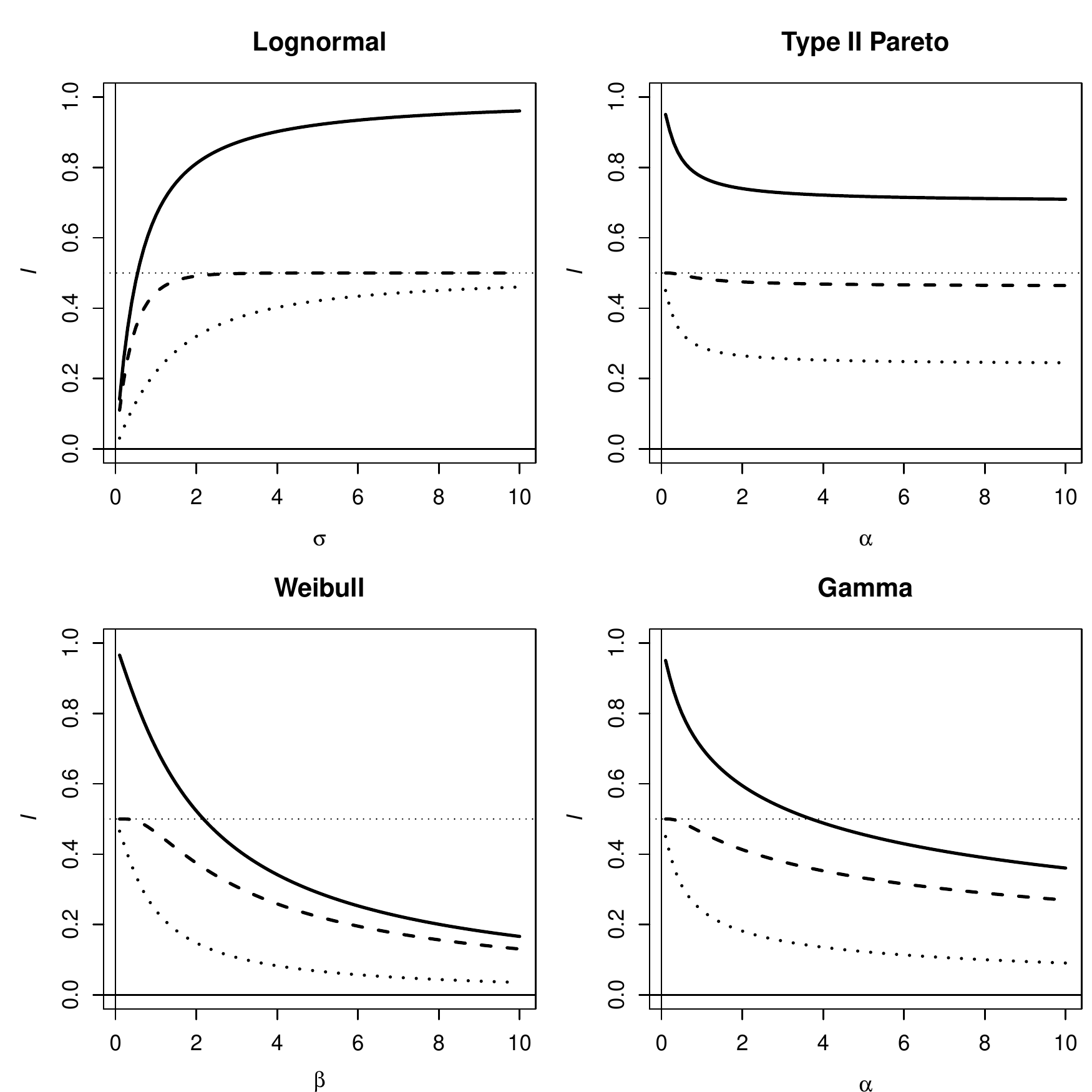}
\end{center}
\end{figure}

\clearpage
\newpage

{\bf Figure 3}

\begin{figure}
\begin{center}
\includegraphics[scale=.9]{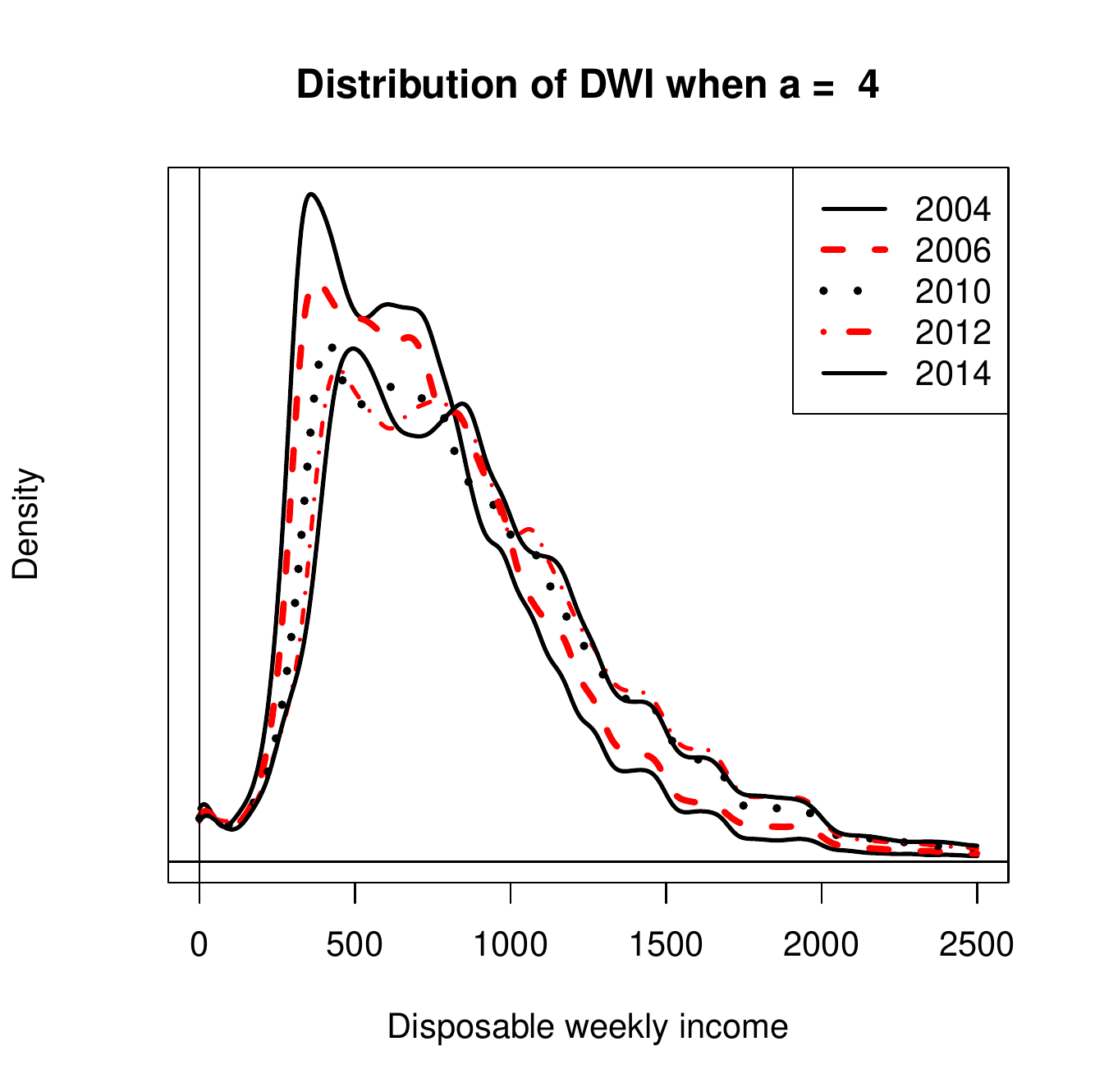}
\end{center}
\end{figure}

\clearpage
\newpage

{\bf Figure 4}

\begin{figure}
\begin{center}
\includegraphics[scale=.9]{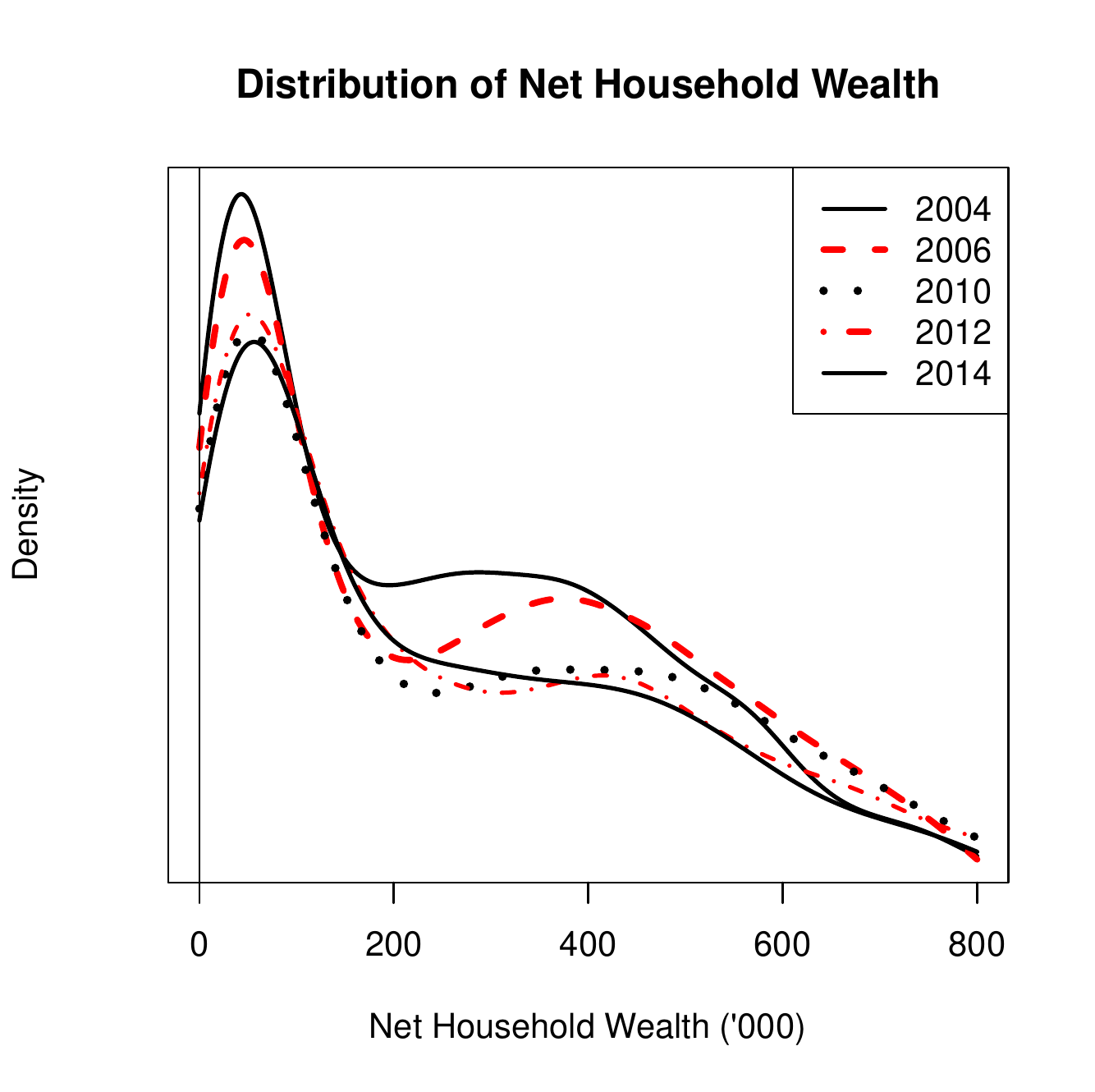}
\end{center}
\end{figure}

\end{document}